\documentstyle[psfig,referee]{mn}
\newif\ifAMStwofonts
\ifoldfss
  \ifCUPmtlplainloaded \else
    \NewTextAlphabet{textbfit} {cmbxti10} {}
    \NewTextAlphabet{textbfss} {cmssbx10} {}
    \NewMathAlphabet{mathbfit} {cmbxti10} {} % for math mode
    \NewMathAlphabet{mathbfss} {cmssbx10} {} %  "   "    "
  \fi
  \ifAMStwofonts
    \ifCUPmtlplainloaded \else
      \NewSymbolFont{upmath} {eurm10}
      \NewSymbolFont{AMSa} {msam10}
      \NewMathSymbol{\upi}     {0}{upmath}{19}
      \NewMathSymbol{\umu}     {0}{upmath}{16}
      \NewMathSymbol{\upartial}{0}{upmath}{40}
      \NewMathSymbol{\leqslant}{3}{AMSa}{36}
      \NewMathSymbol{\geqslant}{3}{AMSa}{3E}

      \let\leq=\leqslant 
      \let\geq=\geqslant 
    \fi
  \fi
\fi % End of OFSS

\ifnfssone
  \newmathalphabet{\mathit}
  \addtoversion{normal}{\mathit}{cmr}{m}{it}
  \addtoversion{bold}{\mathit}{cmr}{bx}{it}
  \newmathalphabet{\mathbfit} % math mode version of \textbfit{..}
  \addtoversion{normal}{\mathbfit}{cmr}{bx}{it}
  \addtoversion{bold}{\mathbfit}{cmr}{bx}{it}
  \newmathalphabet{\mathbfss} % math mode version of \textbfss{..}
  \addtoversion{normal}{\mathbfss}{cmss}{bx}{n}
  \addtoversion{bold}{\mathbfss}{cmss}{bx}{n}
  \newmathalphabet{\mathbfit} % math mode version of \textbfit{..}
  \addtoversion{normal}{\mathbfit}{cmr}{bx}{it}
  \addtoversion{bold}{\mathbfit}{cmr}{bx}{it}
  \newmathalphabet{\mathbfss} % math mode version of \textbfss{..}
  \addtoversion{normal}{\mathbfss}{cmss}{bx}{n}
  \addtoversion{bold}{\mathbfss}{cmss}{bx}{n}     
  \ifAMStwofonts
    \ifCUPmtlplainloaded \else
      \UseAMStwoboldmath
      \makeatletter
      \new@mathgroup\upmath@group
      \define@mathgroup\mv@normal\upmath@group{eur}{m}{n}
      \define@mathgroup\mv@bold\upmath@group{eur}{b}{n}
      \edef\UPM{\hexnumber\upmath@group}
      \new@mathgroup\amsa@group
      \define@mathgroup\mv@normal\amsa@group{msa}{m}{n}
      \define@mathgroup\mv@bold\amsa@group{msa}{m}{n}
      \edef\AMSa{\hexnumber\amsa@group}
      \makeatother
      \mathchardef\upi="0\UPM19
      \mathchardef\umu="0\UPM16
      \mathchardef\upartial="0\UPM40
      \mathchardef\leqslant="3\AMSa36
      \mathchardef\geqslant="3\AMSa3E

      \let\leq=\leqslant 
      \let\geq=\geqslant 
    \fi
  \fi
\fi % End of NFSS release 1

\ifnfsstwo
  \DeclareMathAlphabet{\mathbfit}{OT1}{cmr}{bx}{it}
  \SetMathAlphabet\mathbfit{bold}{OT1}{cmr}{bx}{it}
  \DeclareMathAlphabet{\mathbfss}{OT1}{cmss}{bx}{n}
  \SetMathAlphabet\mathbfss{bold}{OT1}{cmss}{bx}{n}
  \ifAMStwofonts
    \ifCUPmtlplainloaded \else
      \DeclareSymbolFont{UPM}{U}{eur}{m}{n}
      \SetSymbolFont{UPM}{bold}{U}{eur}{b}{n}
      \DeclareSymbolFont{AMSa}{U}{msa}{m}{n}
      \DeclareMathSymbol{\upi}{0}{UPM}{"19}
      \DeclareMathSymbol{\umu}{0}{UPM}{"16}
      \DeclareMathSymbol{\upartial}{0}{UPM}{"40}
      \DeclareMathSymbol{\leqslant}{3}{AMSa}{"36}
      \DeclareMathSymbol{\geqslant}{3}{AMSa}{"3E}

      \let\leq=\leqslant 
      \let\geq=\geqslant 
    \fi
  \fi
\fi % End of NFSS release 2

\ifCUPmtlplainloaded \else
  \ifAMStwofonts \else % If no AMS fonts
    \def\upi{\pi}
    \def\umu{\mu}
    \def\upartial{\partial}
  \fi
\fi

\title{The X-ray emission from  Nova V382 Velorum:\\
I. The hard component observed with BeppoSAX }
\author[M. Orio et al.]
        {M. Orio$^{1,2}$, A. Parmar$^3$, 
R. Benjamin$^4$, L. Amati$^5$, F. Frontera$^5$, J. Greiner,$^6$
\newauthor
H. \"Ogelman$^4$, T. Mineo$^7$, S. Starrfield$^8$, E. Trussoni$^1$\\
$^1$ Osservatorio Astronomico di Torino, Strada
Osservatorio, 20, I-10025 Pino Torinese (TO), Italy\\
 $^2$ Department of Astronomy, 474
N. Charter Str., University of Wisconsin, Madison WI 53706, USA\\
 $^3$ Astrophysics Division, Space Science Dept. of ESA,
   ESTEC, Postbus 299, 2200 AG Noordwijk, The Netherlands\\
   $^4$ Physics Department, 1150
University Avenue, University of Wisconsin, Madison WI 53706, USA\\
$^5$ TESRE-CNR, via Gobetti 101, I-40129 Bologna, Italy\\
$^6$ Astrophysical Institut, 14882 Postdam, An der
Sternwarte 16, FRG\\
$^7$ IFCAI-CNR, via La Malfa 153, I-90146  Palermo, Italy\\
$^8$ Dept of Physics and Astronomy, P.O. Box 87150,
Arizona State University, Tempe AZ  85287-1504, USA}
\date{
      Received; Accepted }

\pagerange{\pageref{firstpage}--\pageref{lastpage}}
\pubyear{1994}
\begin{document}

\maketitle

\label{firstpage}

\begin{abstract}
We present BeppoSAX observations of Nova Velorum 1999 (V382 Vel), done 
in a broad X-ray band covering 0.1-300 keV only 15 days after the discovery 
and again after 6 months.  The nova was detected at day 15 with the BeppoSAX
instruments which measured a flux F$_x\simeq$1.8 $\times  10^{-11}$
erg cm$^{-2}$ s$^{-1}$ in the 0.1-10 keV range
and a 2$\sigma$ upper limit F$_x<$6.7 $\times  10^{-12}$ erg
cm$^{-2}$ s$^{-1}$ in the 15-60 keV range.
We attribute the emission to shocked nebular ejecta
at plasma temperature kT$\simeq$6 keV. At 6 months
no bright component emerged in the 15-60 keV range, but 
a bright central supersoft X-ray source appeared.
 The  hot nebular component previously detected had cooled  to
a plasma temperature kT$<$1 keV.
There was  strong intrinsic absorption of the ejecta in
the first observation and not in the second, because the 
column density of neutral hydrogen
decreased from N(H)$\simeq 1.7 \times  10^{23}$ cm$^{-2}$
to N(H)$\simeq$10$^{21}$ cm$^{-2}$ (close to the the interstellar
value). The {\it unabsorbed} X-ray flux also decreased from
F$_x$=4.3 $\times 10^{-11}$ erg cm$^{-2}$ s$^{-1}$
  to F$_x$$\simeq 10^{-12}$  erg cm$^{-2}$ s$^{-1}$.
\end{abstract}


\section{Introduction}
  {\it Nova Velorum 1999 (V382 Vel)} was discovered
in outburst on 1999 May 22 (Williams \&
Gilmore 1999).  It was the second brightest nova of
this half of the century (V=2.6; Seargent \& Pearce 1999) and
a ``O-Ne-Mg nova'' (Shore et al.
1999a,b). In terms of time to decay from visual maximum
by 2 and 3 mag (see
Della Valle \& Livio 1995) it was a ``fast'' nova, with  t$_2$= 6
d, t$_3$=10 d (Della Valle et al. 1999). The peak ejection velocity
v$_{ej}$ inferred from the emission lines
was v$_{ej}\simeq$4000 km s$^{-1}$ (Shore
et al. 1999a). The estimated distance
is 2 kpc (Della Valle et al. 1999).
 
The nova was declared
a  Target of Opportunity by the BeppoSAX Mission Scientist. The BeppoSAX
X-ray satellite carries instruments that cover the energy range
0.1-300 keV.
 We present results from the coalined Low-Energy Concentrator
Spectrometer (LECS;
0.1--10~keV; Parmar et al. 1997), the Medium-Energy Concentrator
Spectrometer (MECS; 1.8--10~keV; Boella et al. 1997),          
and the Phoswich Detection System (PDS; 15--300~keV; Frontera et al.
1991).
The LECS and MECS consist of grazing incidence
telescopes with imaging gas scintillation proportional counters in
their focal planes. The non-imaging
PDS consists of four independent units arranged in pairs each having a
separate collimator, alternatively
rocked on- and off-source during the observation.
 
 Classical and recurrent novae are expected to emit X-rays in outburst
via three different mechanisms. Luminous {\it ``supersoft''}
X-ray emission of the central source is thought to
indicate that the WD is still burning hyrogen in a shell
(e.g. V1974 Cyg, Krautter et al. 1986, or N LMC 1995, Orio \& Greiner 1999).
This is important because if not all the accreted envelope is
 ejected the WD mass increases after repeated outbursts.
 
Shocks are not the main mechanism 
of nova outbursts: usually the mass outflow is not due
to a ``detonation'' but to a radiatively driven super-wind
 (Bath \& Shaviv 1976).                             
However, {\it shocks  on
a small scale}, due to complex phenomena in the nova wind or in
the interaction between the ejecta and the circumstellar medium, 
are likely to be frequent. The recurrent nova RS Oph, novae V838 Her
(N Her 1991), V351 Pup (N Pup 1991), V1974 Cyg (N Cyg 1992) and
 probably LMC 1992 were   hard X-ray sources shortly after the outburst.
Thermal bremsstrahlung models with temperature in the range 0.5-20 keV  and
luminosity 10$^{33-34}$ erg s$^{-1}$ fit the data.
A list of references includes  Mason et al. (1987), Lloyd
 et al. (1992), Orio et al.
(1996), Balman et al. (1998) and Orio et al. (2000).
This information is largely based on ROSAT
data, and only in the energy range 0.2-2.4 keV. Only V838 Her was
 observed
as early as 9 days after maximum.
There are indications that the plasma temperature might have
been much higher than the ROSAT range (e.g. Lloyd et al. 1992).
 
A third mechanism of X-ray emission is due
to  Compton degradation of radioactive decay,
particularly of $^{22}$Na and $^{26}$Al (Starrfield
 et al. 1992; Livio et al. 1992; Pistinner et al.
1994).  In the range 6-45 keV Compton degradation
might produce significant X-ray flux. Livio et al.         
(1992) argued that in this energy band X-rays emission is expected to
be significant approximately two months  after the outburst. However,
the time for production of X-ray flux is inversely proportional
to v$_{ej}$ and was calculated only for v$_{ej}\leq$1000 km s$^{-1}$,
lower than for V382 Vel. Therefore we do
not rule out that this mechanism may be
relevant at an earlier epoch. Gamma ray observations
of novae  have not yielded very constraining
upper limits so far (see Iyudin et al. 1995 and Wanajo et al. 1999).
The expected gamma ray  luminosity can be as high as
10$^{35}$ erg s$^{-1}$ and the derived X-ray luminosity is at
least two orders of magnitude lower. At the average distance of a galactic
nova it may still be detected with BeppoSAX.
Regardless of v$_{ej}$ and the related time scale for the emission,
the X-ray flux would be higher in the PSD energy range than in the
band covered by the MECS and LECS (see Livio et al. 1992).

%TABLE
\begin{table*}
  \caption{Spectral fit results to the BeppoSAX LECS and MECS
spectrum of Nova Vel observed in June 1999.
 The absorption, N(H), is in units of
 $10^{22}$~atom~cm$^{-2}$. We report the best fit  parameters
and 2$\sigma$ uncertainties for four different models.
 Z is the abundance compared to solar, $\nu$ is the photon index,
F$_x$ is the 0.8--10~keV absorption corrected source flux.
EM is the best-fit emission measure ($\int$n$_e$ n$_H$ dV; the
tabulated values have been divided by 10$^{56}$).}
\begin{tabular}{|l c c c c c c c c |}
\hline
 Model & N(H) &  kT (keV)&  $\nu$  &   Z & Fe &
  F$_x$ (erg cm$^{-2}$ s$^{-1}$)   & EM (cm$^{-3}$) &
  $\chi^2$/dof\\
\hline
 MEKAL & 16.8$\pm$1.0 &  6.1$\pm^{0.9}_{0.7}$
 & & 0.09$\pm^{0.04}_{0.05}$ & & (4.3$\pm^{0.3}_{0.4}$) $\times 10^{-11}$
 & 24.3$\pm$1.8 & 1.2   \\
 VMEKAL & 16.8$\pm 1.1$ & 5.4$\pm^{0.6}_{0.5}$  & & 4$^{+0.5}_{-4.0}$ &
 0.21$^{+0.12}_{-0.09}$ &
(4.6$\pm 0.4$) $\times 10^{-11}$ & 9.9$\pm$0.7 & 1.1 \\
 NEI & 16.7$\pm1.0$ & 6.2$\pm^{1.0}_{0.7}$
 & & 0.09$\pm0.04$ & & (4.3$\pm0.3$) $\times 10^{-11}$ & 26.8$\pm$1.7 & 1.3 \\
POWERLAW & 19.3$\pm^{1.2}_{0.4}$  &  & 2.4$\pm^{0.2}_{0.1}$
 & & & (5.5$\pm^{1.7}_{1.1}$) $\times 10^{-11}$ & & 1.6 \\                     
\hline
\end{tabular}
\end{table*} 
%ENDTABLE
 
In this paper we focus on the evolution of the hard X-ray
emission (range 0.8-60 keV).  No other classical nova had ever
been observed immediately after maximum in such a broad
energy range.  We analyse the 
 supersoft X-ray emission (below 0.8 keV) 
and the related white dwarf atmospheric model in  
a forthcoming Paper II.
\section{The observations}

  V382 Vel was observed for the
first time with BeppoSAX 15 days after the optical maximum, on 1999 June 7-8
for 42.5 ks with the two MECS, for 13.5 ks with the LECS, for
23.3 ks with the PDS. 
The nova  was then observed a second time on 1999 November
23 for 25.9 ks with the MECS,  12.4 ks with the LECS, 12.4
ks with the PDS.
Good data were selected from intervals when the elevation angle above
the  Earth's limb was $>$$4^{\circ}$ and when the instrument
configurations were nominal, using the SAXDAS 2.0.0 data analysis
package.
The standard PDS collimator dwell time of 96~s for each on- and
off-source position was used together with a rocking angle
210\arcmin.
LECS and MECS data were extracted centered on the position of V382 Vel
using radii of 8\arcmin\ and 4\arcmin, respectively.
The background subtraction for the imaging instruments      
was performed using standard files (1997 releases), but is not
critical because the nova
turned out to be a bright source. The background subtraction
for the PDS was
obtained during intervals when the collimator was offset from the
source.
 
In June 1999
 the nova was detected with a count rate 0.1537$\pm$0.0020 cts s$^{-1}$
and 0.0620$\pm$0.0026 cts s $^{-1}$ in the MECS and LECS,
respectively (Orio et al. 1999a).
At higher energies the 2$\sigma$ upper
limits obtained with the PDS were 0.0480 cts s$^{-1}$ in the   15-30 keV band
and 0.0740 cts s$^{-1}$ in the 15-60 keV band.
In the second BeppoSAX observation the count rate measured with the
LECS was extremely high, 3.4860$\pm$0.0021 cts s$^{-1}$,
due to the emergence of the central supersoft X-ray source
 (Orio et al.  1999b). In the range 0.8-10.0 keV
the count rate was only 0.1030$\pm$0.0038. As we mentioned in the introduction,
 in this paper we discuss
 the evolution of the hard  X-ray emission,
the only component detected with the MECS, with a count rate
0.0454$\pm$0.0015 cts s$^{-1}$ (more than a factor 3 lower than in June).
Even in November 2001 there was no PDS detection  with 2$\sigma$ upper limit
0.0800 cts s$^{-1}$ in the 15-50 keV range.
 
For both observations we examined the possibility of variable X-ray flux.
With  Kolmorogov-Smirnov tests we found
 that the flux in the MECS is not variable by more than 15\%
at the 80\% confidence level in the first observation,
and by not more than 40\% in the second observation.
These results are within the statistical fluctuations and do not
imply significant variability.

\section {Spectral analysis and interpretation}
 
We translated the PDS measurements   into upper limits
to the flux assuming a power law spectrum with photon index 2.1
(this result is not critically model dependent). The 2$\sigma$
upper limits obtained are:
 F$_x <$ 6.7 $\times$10$^{-12}$ erg cm$^{-2}$ s$^{-1}$ at 15-60 keV
in June 1999,
F$_x <$ 4.2 $\times$10$^{-12}$ erg cm$^{-2}$
s$^{-1}$ in the 15-50 keV band in November 1999.
\"Ogelman et al. (1987) observed three novae with EXOSAT 3 to 7
months post-maximum: the upper limits for the flux
obtained with  the medium energy (ME) experiment
at 6-50 keV were of the order of 10$^{-11}$ erg cm$^{-2}$.
Our BeppoSAX upper limits are lower  than the flux
measured in the MECS (regardless of the model
assumed for the emission, see below), so we exclude
radioactive decay  as the main mechanism of emission,
because we would measure a lower X-ray flux in the LECS-MECS range
than at 16-45 keV with the PDS (see section 1 and Livio et al. 1992).  
Lacking both a detailed spectral resolution and
models of the evolving nebula predicting its X-ray
luminosity, the observed X-ray flux of novae up to now
 has been fitted with thermal plasma models with
a foreground absorbing column (Lloyd 1992; Balman et al. 1998).
In Table 1 and 2 we show the results of different spectral
fits with models available in the XSPEC
 standard analysis package (Arnaud 1996).
 We fitted thermal models with and without ionization equilibrium
 and a power law model for comparison.
We note that the latter describes the data less
 adequately, consistently with the belief that the X-ray emission is
 due to shocks. The thermal equilibrium
model modified for low
energy absorption, {\sc mekal}, is included to 
compare the data with Mukai and Ishida (2001) and with observations of 
V838 Her and  V1974 Cyg. {\sc vmekal} is used to test the effect of
 varying single element
abundances.  The non-equilibrium model {\sc nei} is  included 
to test departures from ionization equilibrium. These may
be expected when heating (arising from shocks)
and subsequent cooling processes operate on a shorter timescale than the
ionization/recombination times of individual ions. At the energy                range we are studying hydrogen is fully ionized and in
ionization equilibrium (see formula 1e of Rossi et al.,
1997, for the recombination time), however
the metals might be far from equilibrium.
 {\sc nei} is a constant temperature  and  single ionization
parameter model; it is useful in characterizing
the spectrum although it is not
physically detailed. More detailed non-equilibrium models available
in XSPEC
are not specifically suited to the physics of nova shells. 
We find that
 the thermal models yield remarkably similar results,
so they are not well constrained by the data.
 
  The first observation is adequately modeled by a thermal plasma
at a temperature kT$\simeq$6 keV. We compare this
with kT$\simeq$10 keV obtained by Mukai  \& Ishida (2001)
at day 17  post-outburst. 
These authors found that the plasma temperature
cooled again in four subsequent
observations done with RossiXTE.  The models in Table 1 also
indicate  unabsorbed luminosity a few 10$^{34}$
erg s$^{-1}$ (assuming a distance 2 kpc)
and N(H)=0.6-2.7 $\times$ 10$^{23}$ cm$^{-2}$ s$^{-1}$,
which is unusually high. Mukai \& Ishida (2001)
derived a similar N(H) value at day 17 and found that later 
the intrinsic absorption decreased. In our bandpass the
absorption is mainly due to photoelectric opacity of carbon and oxygen
(Morrison \& MacCammon 1983). In the direction of the nova the
 column density of neutral hydrogen is
estimated to be 3.7 $\times$ 10$^{21}$ cm$^{-2}$ (Dickey \& Lockman
1990). Platais et al. (2000) obtained a reddening
consistent with N(H)$\leq$2.5 $\times 10^{21}$ cm$^{-2}$.
The large value of N(H) at the early epoch is therefore due
to {\it intrinsic} absorption of the ejecta, known to be rich
in oxygen and probably carbon as well.
The gas must have been optically thick to lower
energy components in the nebular flux,
such as the supersoft emission of the central source.
Using the simple assumption that EM=$<n_e^2>$ V$_{shock}$, where
n$_e$ is the electron density and V$_{shock}$ is the volume
filled by the shocked mass, the best fit emission measure (EM)
 in Table 1 is consistent with
a shell filled in 15 days with a constant flow at V$_{ej}$=4000 km s$^{-1}$
(V$_{shell}\simeq 6 \times  10^{44}$ cm$^3$)
 and n$_e \simeq 2 \times 10^6$  cm$^{-3}$ (a value
which is on the upper range derived for nova shells).
However,  the shocks could also come  dense clumps along the line
of sight or from a zone which is deeply buried inside the nova shell,
with higher density and much lower volume than the whole nebula.
The two
thermal models {\sc mekal} and {\sc nei} seem to require low abundances for 
       the best fit, however this does not imply low abundance 
Z of {\it all} elements: simply 
in order to match the data, it is necessary to decrease the strength of
the Fe K-$\alpha$ line at 6.97 keV. It is understood using the
{\sc vmekal} model, the only one allowing
specification of single element abundances. Varying other elements,
we derive Fe=0.0-0.39 within the 2$\sigma$ confidence level,
 while 
enhanced abundances are instead perfectly acceptable for the other
elements. The value of $\chi^2$/dof  decreases from 3.1 to 1.2
decreasing Fe from 1 to the best fit value 0.21.
 In Fig. 1 we show (as an example, not
implying it is necessarily the right model) the best {\sc vmekal}
fit (thermal plasma modified by low-energy absorption),
to the combined LECS, MECS and PDS spectrum.
We plot in the figure
the data above 1.8 keV because there is no significant flux
detection below. We compare the residuals of the fit with
reduced abundance and with solar iron. We remind the reader that
the continuum optical depth at energy kT=6.97 keV is negligible,
$\tau_{cont}=0.084 N_{23}$ (where $N_{23}$=N(H)/10$^{23} {\rm           
cm^{-2}}$), and so is the
line optical depth ($\tau_{line}=0.08 T^{-0.5}_{7} n_{e,5}L_{15}$,
where $T_{7}$ is the temperature of the emitting region in units of
$10^{7}$ K, and $L_{15}$  is the
region thickness in units of $10^{15}~{\rm cm}$).
 
 We simultaneously fitted the MECS  spectrum and the LECS  one above
0.8 keV observed in November 1999 with the same models.
Above 0.8 keV, no X-ray emission is expected
from the white dwarf atmosphere: preliminary fits
to the whole 1-10
keV LECS spectrum indicate that we are also dealing with a separate
spectral component.
The plasma temperature is much lower, not exceeding 1 keV,
and the value of N(H) is around the interstellar value,
$\simeq 10^{21}$ cm$^{-2}$. We note that the large intrinsic
absorption of the ejecta almost completely thinned out
and the temperature range is comparable with the one derived for
N Cyg 1992 six months after optical maximum (Balman et al. 1998).
We still derive the best thermal fits with
a low value of the total heavy elements abundance. As
the {\sc vmekal} fit indicates,  at 
 low energy this is not due (or not only) to iron. It could          
be to very unusual abundance ratios of different
elements, or  possibly
 it is an effect of
 lack of ionization equilibrium that {\sc nei} is  not sophisticated enough to
model. However, ionization
equilibrium is commonly assumed after several months from optical maximum
 (see Contini et al.  1995, and references therein).
 
 The LECS-MECS spectrum above 0.8 keV and the best fit with the
{\sc vmekal} model is shown in Fig. 2 for comparison with the first
data set. We found a nebular component even in the
range below 0.8 keV, where the flux from the central source was 
 thought to be dominant. ``Disentangling'' it from the 
atmospheric continuum is  the subject of Paper II. 
 The point we want to make here
is that the shell had significantly cooled and
there was no X-ray emitting plasma at a temperature above 1 keV.
Also, we notice that
the total unabsorbed luminosity
above 0.8 keV from the shocked shell at this stage had decreased
to few 10$^{33}$ erg s$^{-1}$ (approximately one order of
magnitude).
\begin{table*}
\caption{Same entries as in Table 1, for the best fit parameters
to the BeppoSAX MECS and LECS ({\it above} 0.8 keV)
spectrum observed for V382 Vel in November 1999.}
\begin{tabular}{|l c c c c c c c c |}
\hline  
Model & N(H) &  kT (eV)&  $\nu$  &   Z & Fe &
  F$_x$ (erg cm$^{-2}$ s$^{-1}$)   & EM(cm$^{-3}$) &
  $\chi^2$/dof\\
\hline
 MEKAL   & $<$0.18 & 616$\pm 32$ & &
  0.06$\pm^{0.33}_{0.06}$ &  & (9.6$\pm 0.5$)
        $\times 10^{-12}$  & 11.3$\pm$10.0 & 1.2 \\
 VMEKAL & $<$0.21 & 626$^{+100}_{-60}$ & & 0.04$^{+0.10}_{-0.02}$ & 0.04$\pm0.20$ & (9.6$\pm 0.5$)$\times 10^{-12}$ & 10.$\pm$2.0  & 1.2 \\
 NEI     & 0.04$\pm^{0.21}_{0.04}$ & 775$\pm^{45}_{72}$ & &
0.08$\pm^{0.37}_{0.08}$ &  & (9.0$^{1.0}_{2.5}$)
 $\times 10^{-13}$  & 9.9$\pm$2.3 & 1.3    \\
 POWERLAW & 0.28$\pm 0.12$ &   & 5.6$\pm0.4$
 & & & (9.8$ \pm 0.3$) $\times 10^{-13}$  & & 1.2    \\
\hline
\end{tabular}
\end{table*}    
%ENDTABLE
%Figure 1
\begin{figure}
% \vspace{152pt}
\centerline{\psfig{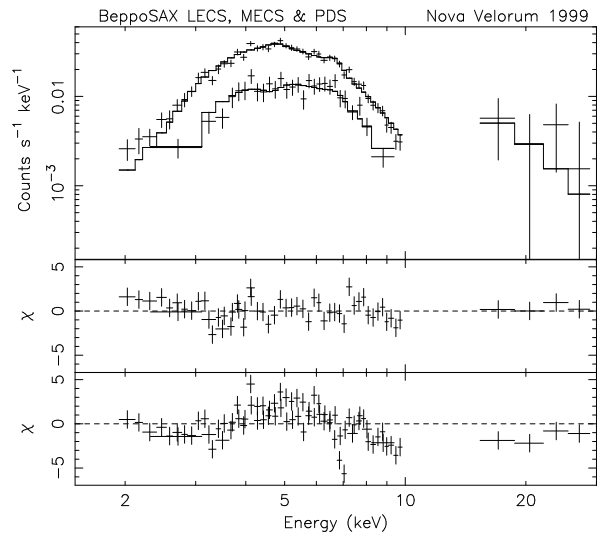}}
\caption{Observation of June 1999:
the LECS, MECS and PDS spectra and the best fit obtained with a {\sc
vmekal} model of a thermal plasma with  depleted iron abundance (see text),
enhanced abundance of all other elements (4 times the solar value),
 kT=6.2 keV N(H)=1.67 $\times$ 10$^{23}$ cm$^{-2}$
 (the reduced $\chi^2$ is 1.13 per 83 dof). The residuals
in units of $\sigma$
are shown in the middle panel; underneath we plot the residuals
of a fit done assuming solar abundances are also shown. }
\end{figure}
%
%Figure 2
\begin{figure}
%\vspace{152pt}
\centerline{\psfig{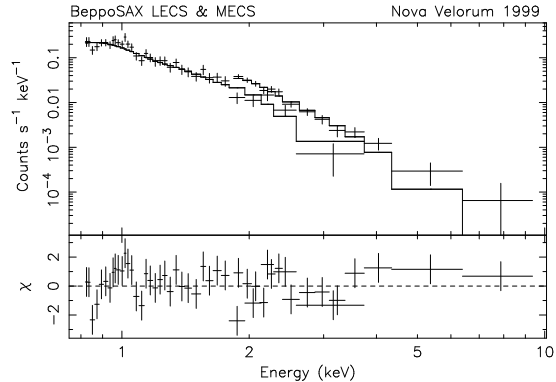}}
\caption{Observed LECS ( only above
0.8 keV), and MECS spectrum  of November 1999 and the best fit obtained
with the {\sc vmekal} model
with kT=626 eV, and N(H)$< 2.1 \times$ 10$^{21}$ cm$^{-2}$ (see Table 2).}
\end{figure}       
\section{Conclusions}

A comparison with the observations published by
Mukai \& Ishida (1999 and 2001) reveals that
Rossi-XTE did not detect the nova on the 3rd day after maximum
in the 2.5-10 keV band
(the upper limit to the flux was 2.5 $\times 10^{-12}$ erg cm$^{-2}$ s$^{-1}$,
see Mukai \& Swank 1999). As we already mentioned,
the nova was observed twice with ASCA shortly after the BeppoSAX
observation in the band 0.4-10 keV. The flux was constant or had increased by
not more than $\simeq$20\% at day 17, and it remained
at approximately constant level until day 59 when
it was observed for the last time with Rossi-XTE
( Mukai \& Ishida, 1999, and 2001).
 Therefore the peak of X-ray
emission occurred during the third week after optical maximum,
a few days after the first BeppoSAX observation,
and a luminosity ``plateau''
followed for at least 39 days. In the  0.8-2.4 keV range the unabsorbed 
X-ray flux was $\simeq 8 \times  10^{-12}$ erg cm$^{-2}$ s$^{-1}$
and  $\simeq 5 \times  10^{-13}$ erg cm$^{-2}$ s$^{-1}$ in the first
and second BeppoSAX observation respectively.
These fluxes can be compared with the ones derived with ROSAT: for 
of V1974 Cyg the flux in the 1-2.4 keV range 
reached a few 10 $^{-11}$ erg  cm$^{-2}$ s$^{-1}$ at its peak around
 day 150 post-maximum (Balman et al. 1998). For V351 Pup 
F$_x \geq 10 ^{-12}$ erg  cm$^{-2}$ s$^{-1}$ in the band 
0.8-2.4 keV 16 months post-maximum (see Orio et al. 1996).
 These two nova shells must have been intrinsically
more luminous in these energy ranges and for a longer time, however
no comparison is possible at energy above 2.4 keV.  

 From our observations, and from the comparison
with the ASCA and RossiXTE light curve derived between them by
Mukai and Ishida (2001), we concluded the following points:
               
   \begin{enumerate}
 
\item We attribute the hard X-ray
emission to shocks in a small portion of the the ejected nebula,
in agreement with Mukai and Ishida (2001). Comptonized
X-rays  from radioactive decays are not definitely
ruled out but are not the main
source of the hard X-ray flux detected in these observations;
 
 \item The unabsorbed X-ray luminosity in the range above 0.8 keV was
few times 10$^{34}$ erg s$^{-1}$ 15 days post-maximum. 
After a period of constant level, it decreased to 
few times 10$^{33}$ erg s$^{-1}$ five and a half months later;
 
\item In the first observation large intrinsic absorption of the ejecta
prevented detection of X-ray flux below 2 keV, while the ejecta
were transparent to supersoft X-ray radiation in the second observation;
 
 \item
The observed X-ray emitting nebular plasma was at a temperature
in the several keV range at 15 days, it reached
10 keV at day 17 and immediately
started cooling. It cooled to a temperature
below 1 keV at 6 months;
      
 \item There is a high probability that the shocked material was significantly
depleted in iron and that abundance ratios of different
elements were quite peculiar, although we cannot be more specific.
While Mukai and Ishida (2001) dismiss the derived iron abundance 
as due to non-availability of completely adequate models,
we suggest that it might actually be real;

\item A comparison with ROSAT observations of V1974 Cyg 
and V838 Pup shows that the
evolution and cooling of the hard X-ray component 
from nova nebulae occurs on quite different time scales.
   \end{enumerate}

\section*{Acknowledgments}
 
 We thank L. Piro, BeppoSAX Mission Scientist, for
support; we are grateful also to M. Della Valle for useful
discussions,  to C. Markwardt for computer support,
and to K. Mukai who shared the ASCA
results with us. M. Orio has been supported by ASI and
by the UW College of L\&S;
 J. Greiner by the German Bundesministerium f\"ur
Bildung, Wissenschaft, Forschung und Technologie (BMBF/DLR) under
contract No. 50 QQ 9602 3.

\end{document}